\documentclass[journal]{IEEEtran}
\usepackage{CJK}
\usepackage{amssymb}
\usepackage{amsmath}
\usepackage{graphicx, subfigure}
\usepackage{url, cite}
\usepackage{verbatim}
\usepackage{booktabs}
\usepackage{multirow}
\usepackage{bigstrut}
\usepackage{booktabs}

\usepackage{subfigure}
\usepackage{graphicx}

\usepackage{amsmath,cite,graphicx,times,subfigure,url,verbatim}
\usepackage{algorithmic}
\usepackage{algorithm}
\usepackage{epstopdf}
\usepackage{textcomp}

\usepackage{tabularx}

\newtheorem{subsec:coding}{subsec:coding}

\usepackage{color}

\begin{document}

\title{CreativeBioMan: Brain and Body Wearable Computing based Creative Gaming System}

\author{
Min~Chen,~Yingying~Jiang,~Yong~Cao,~Albert~Y.~Zomaya

\thanks{M. Chen, Y. Jiang and Y. Cao are with Huazhong University of Science and Technology, China.

A. Y. Zomaya is with The University of Sydney, Australia}
}

\markboth{}{}

\maketitle

\begin{abstract}
Current artificial intelligence (AI) technology is mainly used in rational work such as computation and logical analysis. How to make the machine as aesthetic and creative as humans has gradually gained attention. This paper presents a creative game system (i.e., CreativeBioMan) for the first time. It combines brain wave data and multimodal emotion data, and then uses an AI algorithm for intelligent decision fusion, which can be used in artistic creation, aiming at separating the artist from repeated labor creation. To imitate the process of humans¡¯ artistic creation, the creation process of the algorithm is related to artists¡¯ previous artworks and their emotion. EEG data is used to analyze the style of artists and then match them with a style from a data set of historical works. Then, universal AI algorithms are combined with the unique creativity of each artist that evolve into a personalized creation algorithm. According to the results of cloud emotion recognition, the color of the artworks is corrected so that the artist¡¯s emotions are fully reflected in the works, and thus novel works of art are created. This allows the machine to integrate the understanding of past art and emotions with the ability to create new art forms, in the same manner as humans. This paper introduces the system architecture of CreativeBioMan from two aspects: data collection of the brain and body wearable devices, as well as the intelligent decision-making fusion of models. A Testbed platform is built for an experiment and the creativity of the works produced by the system is analyzed.
\end{abstract}


\section{Introduction}

Machines have surpassed humans in many aspects with the development of AI technology. Nevertheless, current AI technology is mainly applied in rational fields of work. The computing and logical analysis abilities of AI are especially better than those of humans. People usually think that the appreciation and creation of beauty is an exclusive peculiarity of humans. However, AI in rapid development has shown its specific aesthetic perception and creativity while imitating the skills of humans. Current AI is not only a program that can compute, but it can also generate creative digital images. For example, in~\cite{BJoshi}, the authors apply neural style transfer to redraw key scenes in Come Swim in the style of the impressionistic painting that inspired the film. There also exists some works to apply AI in fashion. For example, the work of Shuhui Jiang et al.~\cite{SJiang} can automatically generate a clothing image with the certain style in real time with a neural fashion style generator. In paper~\cite{SZhu}, the authors present a novel approach for generating new clothing on a wearer through generative adversarial learning while at the same time keeping the wearer and her/his pose unchanged. To solve the problem of visual creation process and let non-professionals participate in artistic creation, brain waves and multimodal emotion data offer new trains of thought in the art creation process.

Cognitive biometrics has attracted increasing attention. EEG and ECG signal measurement can detect the reaction of the brain and record some electromagnetic, and untouchable neural oscillations~\cite{TAlotaiby}. Emotion data adds more personalized elements to art creation, endows a soul to the created work. The AI algorithm can be used to process the style of artistic contents created by the artist according to a combination of the brain wave of the artist and their emotion. Thus, the efficiency of the art creation process will be increased greatly if an artist can be emancipated from repetitive labors.

On this basis, we put forward CreativeBioMan for the first time to jointly encode brain wave and multimodal emotion data, and apply it to creative games. The robot has creativity like humans to draw paintings and create some artworks. Creative means a creative algorithm. BioMan means a humanoid biological robot, which can collect the user's biological signals and design the algorithm according to the human cognitive process; it is a coalition of machine and human along with a virtual and creative AI entity.

Compared with the above existing work, this paper not only innovatively uses EEG data for artistic creation, but also incorporates user emotions into the works, which enriches the meaning of the works. In addition, this paper did a systematic integration work and build a complete creative system platform. In short, our main contributions are as follows:

\begin{itemize}
  \item This paper puts forward the creative game system, CreativeBioMan, and for the first time develops a creative game using brain wave and multimodal emotion data, completes decision fusion and superposition with an AI algorithm, and makes their respective advantages complementary to each other into something that evolves into a personalized creative algorithm.
  \item This paper introduces the two important parts of CreativeBioMan in detail. Brain-wearable devices and wearable clothes were used by an emotional robot to respectively collect brain wave data and multimodal emotion data. Styles were then matched based on motor imagery model to help artists to make their creation.
  \item This paper describes the building of the CreativeBioMan experimental platform. In addition, to verify the creativity of system, an experiment was designed for an expert to analyze the fidelity of the artworks generated by the machine compared to artworks created by humans.
\end{itemize}

The remainder of this paper is structured as follows: Section II introduces work on AI related to creativity. Section III discusses the data collection, intelligent decision fusion, and artistic creation in the creative game system CreativeBioMan. Section ¢ô expounds the motor imagery model, VGG-19 network model, and attention-based RNN emotion recognition model used in this paper. Section V introduces the testbed in the experiment and analyzes the fidelity of the works of CreativeBioMan and the creativity of the system.

\section{Related work}  \label{sec:arch}
Turing put forward the question, ``Can machines think''? in 1950. Turing created the ``imitation game'' to test whether computers could make a human believe it is a person. In the same way, the concept of the imitation game can be applied to the argument ``Can computers have creativity''? Related work is mainly divided into two parts: one is the feasibility of machine application in different fields to create; the other is to summarize methods used for machine creation.

First, in 2014, the Google DeepMind team developed the very successful creative computer program, AlphaGo. At present, it is the most powerful Go player in the world. This computer uses deep thinking to teach itself how to succeed in a Go game.

In the South by Southwest (SXSW) conference held in 2014, IBM cooperated with the Institute of Culinary Education (ICE) to use the recipe system of Watson to build a mobile canteen with the name ``Cognitive Cooking''~\cite{FPinel}. Watson used more than 30,000 recipes from Wikipedia or an ICE exclusive database. The chemical composition and flavor of almost all the food materials could be known, and a database that contained different cultural flavor preferences was used to suggest different dishes. The mobile canteen of Watson could not only provide novel catering but also assess its products in detail with an advanced feedback mechanism.

In 2016, the Logojoy company brought about a completely new mode of logo design~\cite{BAStanescu}. The visual effect of a logo created by their program is almost same as that of a digital image made by a human. To create a logo, the name of the company is first needed. Then, at least five icons at are selected to give ``inspiration'' to the computer for estimating preferences. Next, a color palette is chosen to guide the selection of logo color. Finally, a selective procedure is given. Five icons at most can be selected to be contained in the logo. A slogan for the company can be added in if desired. Then, the algorithm is used to process the information collected and the result is displayed on the screen.

The obvious different features of these computers are ``deep learning'' and ``neural network''. The computer program that contains deep learning, a neural network, and a back-propagation algorithm can find the complex mode of large-scale data integration to guide the machine to set new parameters and compute later data~\cite{YLeCun}. These computer algorithms are different from the organon theory that a computer ascertains results first. A deep learning computer sets the initial guide. Therefore, the computer can analyze data at the beginning to generate a new guide or parameters and recognize potential new results. The complexity of the neural network and dynamic pattern recognition can allow the creative thinking of the computer to be able to generate a novel concept for a field. At present, almost all the literature related to creative computers is outside of the field of art education.

On the other hand, as for about the research methods, Fink et al. collected EEG signals of professional dancers during improvisational dancing. They observed evidence that during the generation of alternative uses professional dancers show stronger alpha synchronization in posterior parietal brain regions than novice dancers ~\cite{AFink}. Fortino et al.~\cite{WTFreeman}-~\cite{GFortino} proposes a completely new architecture that support the development of novel smart wearable systems for cyberphysical pervasive computing environments. And in~\cite{AJChampandard}, the authors introduce an augmented CNN architecture that bridges the gap between generative algorithms and pixel labeling neural networks. Leon et al.~\cite{LAGatys} uses neural representations separate and recombine content and style of arbitrary images, providing a neural algorithm for the creation of artistic images.


\section{System Architecture}  \label{sec:tech}
\begin{figure}
\centerline{\includegraphics[width=3.5in]{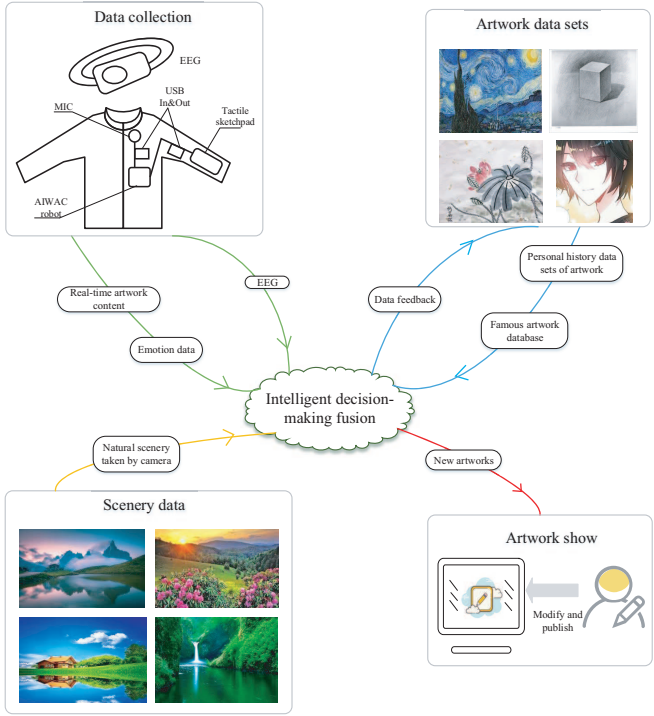}}
\caption{system architecture.}
\label{Fig1}
\end{figure}

The system architecture includes five modules: user data acquisition module, historical creation data set, landscape data acquisition module, cloud processing module and artwork publishing module. Firstly, in the process of system design, we use an emotional robot of brain-wearable devices and wearable clothes to collect data. Secondly, we need to collect different styles of paintings created by users and use convolutional neural network to extract the style features of paintings. The third part is the landscape data acquisition module. When users are outdoors and do not want to create, they can directly use the camera to take pictures and the system can extract the content features of the artworks. Then, an artificial intelligence algorithm is deployed in the cloud. After completing the intelligent decision fusion and creating artwork in the cloud, the results are sent to the intelligent terminal and displayed. The system architecture is shown in Fig.\ref{Fig1}.


\subsection{Data collection of brain and body wearable devices}

£¨1£©EEG data

In the CreativeBioMan creativity game, we use the independently designed brain-wearable device to collect the brain wave data of users. collected. In research findings, different neural activity generates a different brain wave pattern~\cite{ADietrich}. For example, when a human is idle or in a state of artistic creation, then the brain wave is a theta wave at a frequency between 4-8 HZ.

£¨2£©Real-time artistic creation

As for the drafting of real-time artistic works created by artists, it is called a content feature data of artworks in this paper. The drawing board configured in the wearable clothing emotional robot can be used to collect data in real time. Then, the image data can be transmitted to the cloud in real time. The wearable clothing emotional robot is a good tool for recording the creativity of artists who have divergent thinking, are good at improvisation, and want to capture inspiration at a certain moment. When artists travel and are inspired to paint a picture but there are no painting supplies available, the wearable clothing emotional robot can complete the creation for them.

£¨3£©Multimodal emotion data

An AIWAC BOX is an embedded hardware product independently developed by our laboratory. It can be used for real-time voice interaction with users, detection of users' emotional information, and analysis of users' emotional state based on artificial intelligence algorithms~\cite{MChen2017}. And it is configured in the wearable clothing emotional robot as the hardware core of emotional recognition and interaction. It is mainly used to collect the artists' multimodal emotion data and upload data to the cloud. The peripheral modules of the wearable clothing emotional robot include a communication module, a camera module for image collection, MIC, and a playing module related to voice data collection and interaction. The wearable clothing robot that integrates the AIWAC Box has nine personality characteristics: courage, prudence, sincerity, virtuousness, confidence, modesty, tenacity, foresight, and optimism. It can recognize 21 different emotions of humans. Emotion is the soul of an artwork. After the multimodal data collected by wearable clothing emotional robot is recognized and analyzed in the cloud, the artists' emotion is endowed to the artwork by adjusting its lines and colors.

£¨4£©Artwork data sets

The artwork data sets includes a personal historical work data set for each artist. Besides the content feature data, brain wave data, and multimodal emotion data of brain-wearable devices and the wearable clothing emotional robot that can serve as the data source for artistic creations of machines, an original artwork data set in a large quantity can also be input in the system. The data sets are classified according to the different creation styles of artists. After the motor imagery model is used to classify the style features of EEG data, they match the style features of the historical works data set~\cite{LAGatys}. In this way, the styles of the content features can be transferred to create artworks with specific style and contents. The algorithm learns and is trained by combining the historical data set, EEG data, and emotion data. The system can generate paintings in specific emotional themes and styles. This is a digital game and a way of presenting the creativity of AI. Artwork data sets integrate the data of the famous artists, which forms a rich database of works. It can recommend the style features for ordinary users and match the style of the user's EEG sports imagination to extract the style features.

\begin{figure*}
\centerline{\includegraphics[width=6.5in]{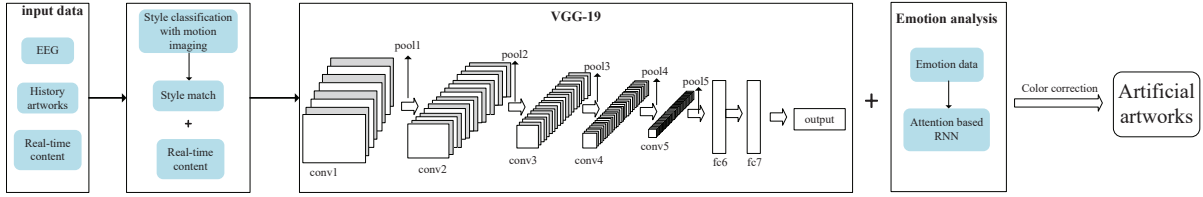}}
\caption{Algorithm flowchart.}
\label{Fig2}
\end{figure*}

\subsection{Intelligent decision fusion and creative game production}

It is necessary to rapidly transmit the data to the cloud for intelligent decision fusion after obtaining EEG data, the content feature data of real-time artistic creation, and multimodal emotion data. Firstly, the cloud uses a motor imagery algorithm to classify the style features of EEG data and analyze the styles desired by the artists. In this paper, style features are divided into four classes: oil painting, traditional Chinese painting, sketch, and cartoon. Then, they are matched with the style features of the historical artworks that have been uploaded to the cloud. After the style of EEG data is matched to that of the corresponding historical data set, the system confirms the actual style of the artistic creation. Content features are determined by the work draft created by the artists in real time.

To form an artwork with specific contents and styles, the VGG19 network algorithm is deployed in the cloud extracts and rebuilds content and style features. In addition, the AI algorithm in the cloud includes an attention-based Recurrent Neural Network (RNN) algorithm for recognizing and analyzing emotion data. The RNN algorithm can effectively memorize relevant feature information of context. By introducing an attention mechanism into RNN algorithm framework, a new weight pooling strategy can be introduced in the network to protrude the part of the voice that has intense emotional features. After the emotion of the artists is recognized in the cloud, the artworks are rectified with contents and style. The change of lines and color are used to express the artists' state of mind when creating artworks.

\section{THE ALGORITHM MODEL}  \label{sec:design}
The creativity of the CreativeBioMan system is determined by the performance of the AI algorithm deployed in the cloud. In this paper, a motor imagery model is used to classify the styles of EEG data, and then a VGG-19 network is used to rebuild style and contents to create a new artwork. An attention-based RNN algorithm is used to raise the accuracy rate of emotion recognition to analyze and recognize emotion data. The system rectifies the colors and lines of artworks that are created according to the emotion recognition result and fuses the emotion into the works. The systematical algorithm procedures are shown in Fig.\ref{Fig2}.

\subsection{EEG data processing and motor imagery model}
Common spatial pattern(CSP) is a frequently-used method in brain-computer interface (BCI) research based on EEG. A data set is required to have a label. The class is known in each experiment. In the task of brain signal classification, it is known that the data collected in a single experiment is a matrix of size $N\times P$, noted as $E_i$. N is the number of channels for signal collection, $P$ is the number of samples of single channel, and $i$ signifies the ith class. If there are $M$ experiments in the $i$th class, then there will be $M$ $E_i$ matrices. Being different from the traditional mean normalization method for obtaining the covariance of the class, the $M$ $E_i$ matrixes in the same class are linked in the direction of the row vector to obtain the entire EEG signal data $T_i$ in the $i$th class in size $N\times(M\times P)$. Then, the corresponding space covariance is obtained according to the $T_i$ matrix in the $i$th class, as shown in Formula\eqref{eqC}.

\begin{equation} \label{eqC}
C_i=\frac{T_{i}T_{i}^{T}}{tr(T_{i}T_{i}^{T})}
\end{equation}

The $i\in \{1,2\}$, CSP method is used to obtain the space filtering matrix $W$ of the two classes and validate Formulas \eqref{eqW1} and \eqref{eqW2}.

\begin{align}
W^{T}C_{1}W &=\Lambda_{1} \label{eqW1} \\
W^{T}C_{2}W&=\Lambda_{2} \label{eqW2}
\end{align}

The artwork style in this paper is a four-class problem. A 1-to-1 strategy is chosen to build the CSP in multiple classes. The four classes are combined anew in pairs and six space filtering matrixes $W$  will be obtained. The best six column vectors are chosen for each space filtering matrix. Each column vector can be seen as a filter. There are $6\times6=36$ column vectors in total. Therefore, $36 \times N$ mixed space filtering matrix $\overline{W}$ can finally be obtained. It is notable that $\overline{W}$ shall be saved. In the test, the testing data directly uses the mixed space filtering matrix obtained from set training for filtering. The mixed space filter $\overline{W}$ is used to finally filter the EEG data $E_i$ in a single experiment and obtain $X_i$ of $36\times T$, as shown in Formula\eqref{eqX}.

\begin{equation} \label{eqX}
X_i=\overline{W}E_i
\end{equation}

Then, the features of signal $X_i$ are extracted after space filtering. Firstly, the variances of the row vectors of $X_i$ are obtained. Because of the difference among single EEG signals, the difference among some of the eigen values is large. Therefore, the logarithm of variance is used to alleviate the difference among data, as shown in Formula\eqref{eqv}.

\begin{equation} \label{eqv}
v_i=log(var(X_i))
\end{equation}

$v_i$ is the eigen vector of a single experiment $E_i$. There are six eigen elements in total. In other words, each sample contains six eigen values. Then, LSTM is used to train and build a classifier.

\subsection{Artwork contents and style processing model based on VGG-19}
The processing of artworks consists of extracting the particular creation style of the artworks. The works in the same class are extended, the contents of the works are extracted, and then artworks are created. A VGG19 network is used in this paper to extract and rebuild the features of styles and contents, involving 16 convolutional layers and five pooling layers. As for the extraction of content features, a convolutional neural network of five layers is used for convolution. Mean pooling is operated after each convolutional layer and a content feature matrix is finally generated. As for the extraction of style features, first all the feature maps at a certain layer are processed after being put in the network. There are feature maps in a large quantity at each layer. Then, the inner product of each pair of the feature maps and the style feature matrix that contains the texture and color information of the maps are obtained. After content features and style features are extracted, artistic paintings are created. Style features, content features, and white noise images are input in the VGG19 network. The gradient descent method is used to solve Total Loss function, such as the minimum value of Formula\eqref{eqL}. The output result of the white noise image is used to update $x$ constantly and use the VGG19 network to rectify the result. For the goal of decreasing a total loss, a painting work based on a creation of the artist can finally be obtained.
\newcommand{\myvec}[1]%
{\stackrel{\raisebox{-2pt}[0pt][0pt]{\small$\rightharpoonup$}}{#1}}
\begin{equation} \label{eqL}
L_{total}(\myvec{p},\myvec{a},\myvec{x})=\alpha L_{content}(\myvec{p},\myvec{x})+\beta L_{style}(\myvec{a},\myvec{x})
\end{equation}

$L_{content}(\myvec{p},\myvec{x})$ is the loss of contents; $L_{style}(\myvec{a},\myvec{x})$ is the loss of style; $\alpha$,$\beta$ is the factor of influence.

\subsection{Attention-based RNN emotion analysis model}
In this paper, we use attention-based RNN model to evaluate a human's emotion, and the higher of the similarity between the current input and the target state, the greater of the weight will be assigned to the current input. Besides, The softmax function is introduced here to calculate the parameter $\alpha_t$ output, which is defined to be as:

\begin{equation} \label{eqalpha}
\alpha_t=\frac{exp(\mu^Ty_t)}{\sum_{t=1}^{T}exp(\mu^Ty_t)}
\end{equation}

Therefore, the output of  the attention model is:

\begin{equation} \label{eqz}
z=\sum_{t=1}^{T}\alpha_ty_t
\end{equation}

A different wave signal in the time domain corresponds with a different weight. In an area with concentrated emotional information, $\alpha_t$ is large; in a blank frame or in an area without emotional information, $\alpha_t$ is small. Basic acoustic features are mapped to be discrete emotion feature labels through the pooling layer and softmax through RNN and attention computing.

\section{Testbed and Experiment}\label{sec.conclusion}

\subsection{System testbed}
\begin{figure}
\centerline{\includegraphics[width=3.3in]{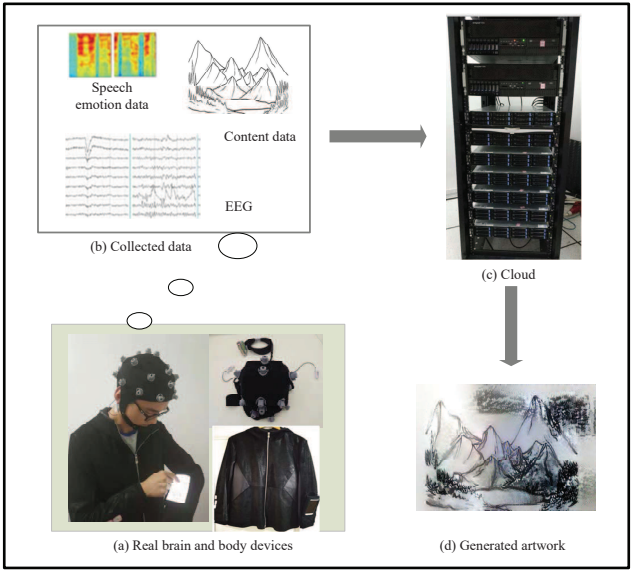}}
\caption{System testbed.}
\label{Fig3}
\end{figure}

We built a testbed for the CreativeBioMan system, including brain-wearable devices, a wearable clothing emotion robot, and a data center in the cloud, as shown in Fig.\ref{Fig3}. For brain wave signal collection, we chose the ADS1299-8 chip of the TI company as the chip and CH559L was the main control chip. The wearable clothing emotion robot integrated AIWAC BOX, the intelligent drawing board, and the MIC voice collection module. Wireless communication was used for data communication with the whole system. The Inspur Big Data Center in the cloud was equipped with two management nodes and seven data nodes. Data of 253 TB could be saved on it, which offered a sufficient hardware guarantee to the real-time computing and analysis of the AI algorithm.

\begin{figure*}[t]
\centering
\subfigure[Artworks of CreativeBioMan system.]{

\includegraphics[height=2in,width=2.2 in]{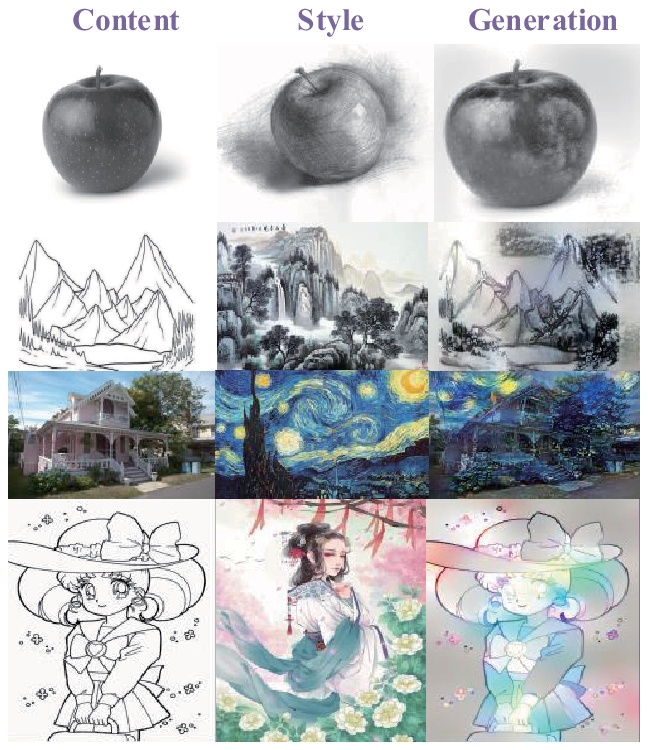}
}%
\subfigure[Verisimilitude Rate of CreativeBioMan system.]{
\includegraphics[height=2.15 in, width=2.3 in]{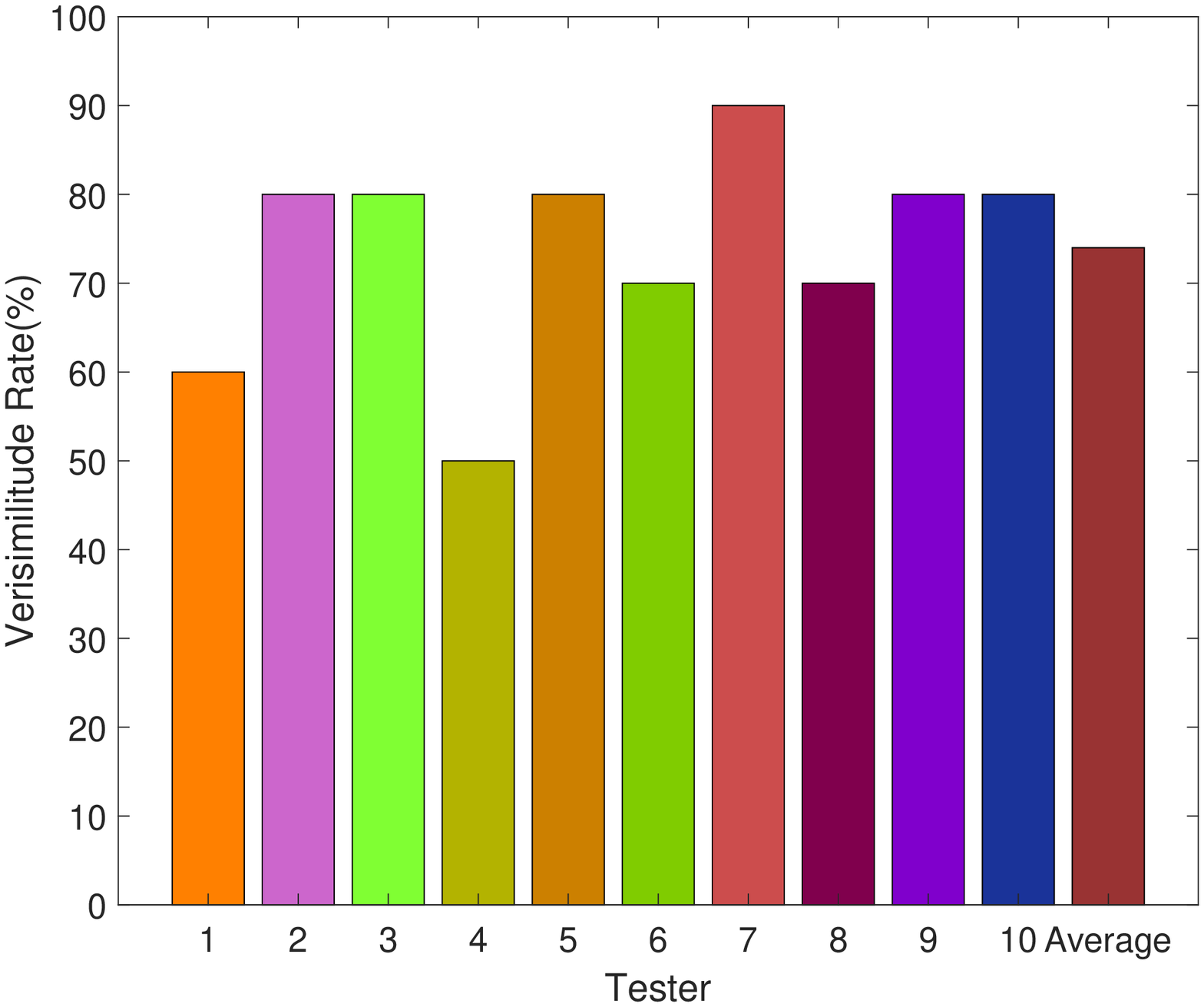}
}%
\subfigure[Delay of CreativeBioMan System.]{

\includegraphics[height=2in,width=2.3 in]{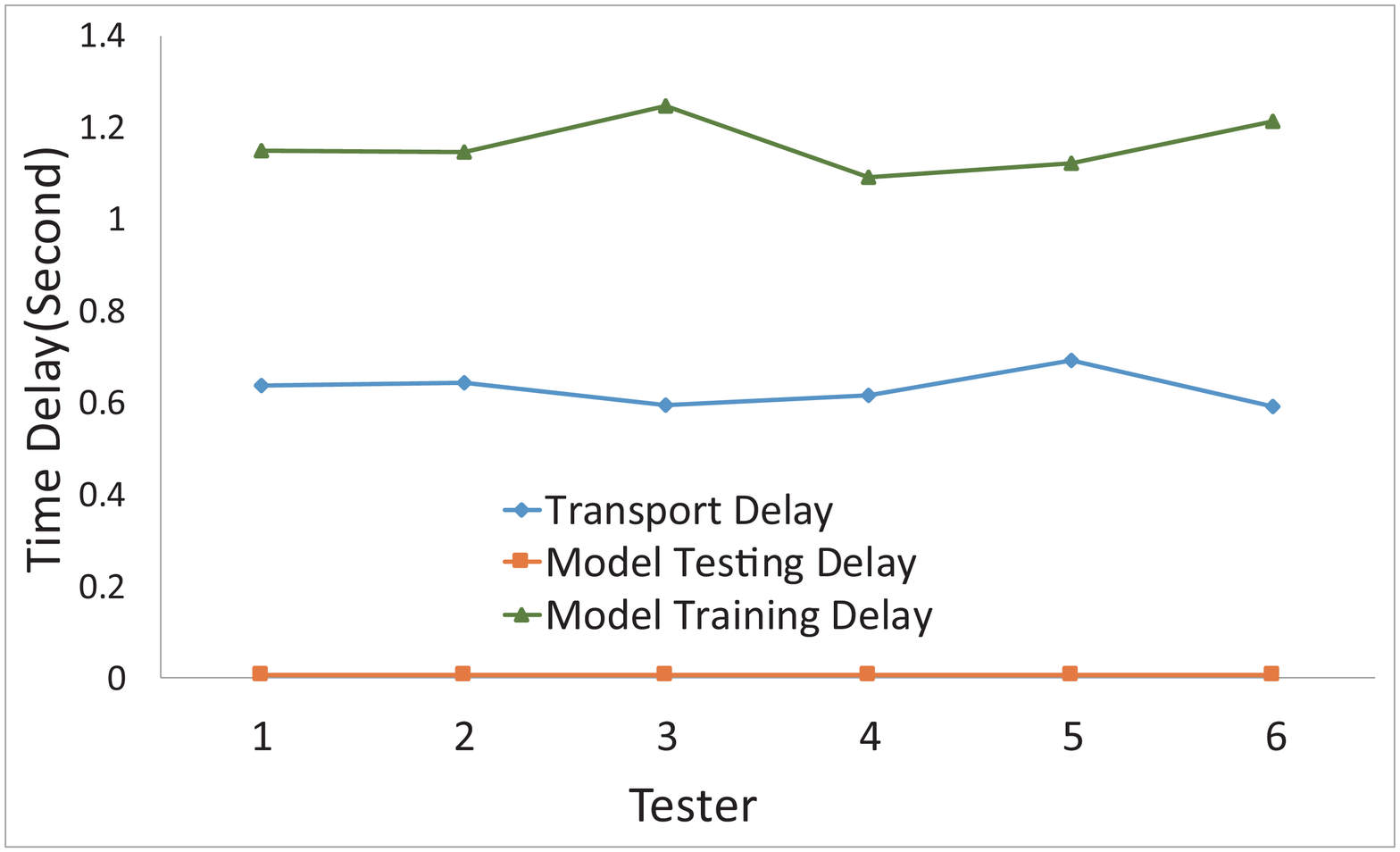}
}%
\caption{Experiment results}
\label{Fig6}
\end{figure*}

The EEG signal sampling frequency was 512 Hz. The EEG signal was collected in three creation processes for each test subject. At the same time, users' emotion was recognized using the wearable clothing emotion robot. After obtaining users' emotion data, the hue of the final works could be slightly adjusted according to the users¡¯ emotion. If the users' emotion was positive, then the hue of the work was set to be warm; if their emotion was to the contrary, then it would be set to be cold. The duration of each monitoring last for 8-15 minutes.

After each test, users need to label the signal of their whole creation process to mark the style of painting that they want. Data The data set is was used to train the imagine classification model for different painting styles to match the styles of painting in the users¡¯ historical works. The data sets built are built were used to train the users' personalized creation process creation process.

First, we preprocessed the 22 paths of EEG signal for each user. A 5-order low-pass Butter worth filter with $fz = 50$ Hz was used to filter the radio-frequency component. The EEG signal was divided into frames and the size of a window was 256. Short-time discrete Fourier transformation was used to extract the rhythm bands of the EEG signal, and then respectively obtain the energy value of $\delta$¡¢$\theta$¡¢$\alpha$, $\beta$, approximate entropy, the largest Lyapunov index, and the Kolmogorov entropy as the signal features of the EEG. They were input into the LSTM network for classification and to obtain the corresponding labeling result.

We collected between five to ten works of users in different styles and marked the painting style of each work. The labels were divided into four classes: oil painting, traditional Chinese painting, sketch, and cartoon. Histogram equalization was used to obtain the luminance image in the whole luminance range. All users¡¯ historical works were resized to be the same size and then they were uploaded to the server.

\subsection{Experimental Results and Analysis}

By establishing the above platform and data set, the CreativeBioMan system could generate paintings with the users' style. This paper assessed the system according to the picture effect generated by the system and the fidelity of artworks.

The artworks generated from users' data set by the style transfer model are shown in Fig.\ref{Fig6}(a). The left shows the painting work created by the users in real time, the middle is the corresponding work obtained according to the labeling value of the EEG signal, and the right side is the mature painting work generated.

In the definition of the creation fidelity of artworks, the works generated by the computer were mixed with those created by real artists and other painters to distinguish and select them. If the painters could not select the works created by the computer, it indicated that the computer had a similar creativity to that of real painters, that is, the fidelity of the creation was very high. Concretely, it can be expressed by a mathematical formula, as follows:

\begin{equation}
life\_like=\frac{\sum_{i=1}^n\sum_{j=1}^mgoal_{i,j}}{n \times m}\times non\_machine \times 100\%
\end{equation}

In the formula, $goal\_{i,j}$ equals 0 if the $i$th judge finds the work created by computer in the $j$th test set; otherwise, it equals 1. In addition, $m$ means $m$ test sets, $n$ means $n$ judges, and $non\_machine$ is the proportion of painting works in the test set which were not created by the computer. The experimental results are shown in Fig.\ref{Fig6}. From Fig.\ref{Fig6}(b)£¬we can see that in the test on the artworks of the nine test subjects, the fidelity rate of the creation of the system corresponding to eight test subjects was higher than 50\%. The fidelity rate corresponding to No. 1 and No. 3 test subjects were 85\%. This indicates that the creativity of the system we built is high. In addition, we also test the running time delay of the system and the result is shown in \ref{Fig6}(c). We can see that the data transmission delay is 0.6-0.8 second, the training time model is 1-1.3 second. Model test time is less than 50ms. Therefore, deploying the algorithm in practical application is feasible.


\section{Conclusion}\label{sec.conclusion}
The AI in rapid development is not limited to use in computing and logical analysis. To endow aesthetic judgment and creativity of human to machine, the creative game system, CreativeBioMan, was put forward in this paper. Brain-wearable device and a wearable clothing emotion robot were used to respectively collect the EEG data and multimodal emotion data during the artists' creation. By combining the artists' previous artworks and using the AI algorithm in the cloud for decision fusion, artists were helped with artwork creation. The AI algorithm model was introduced in detail in this paper, including EEG data processing, style classification based on motor imagery model, style reconstruction and content reconstruction model based on VGG-19 network, and the attention-based RNN emotion recognition model. Finally, a testbed platform of the creative game was built in this paper to analyze the fidelity rate of the works generated by the system and its creativity in detail. In the future work, we consider using EEG data to record the brain state of the user's creation, read out the brain's awareness, and then create a more intelligent and creative game system.

\section*{Acknowlegements}\label{sec.acknowlegements}
This work was supported by the National Key R\&D Program of China (2018YFC1314605, 2017YFE0123600).

\bibliographystyle{IEEEtran}


\begin{thebibliography}{99}

\bibitem{BJoshi}
B. Joshi, K. Stewart, D. Shapiro, ``Bringing Impressionism to Life with Neural Style Transfer in Come Swim'', Proceedings of the ACM SIGGRAPH Digital Production Symposium. ACM, 2017, pp. 5.

\bibitem{SJiang}
S. Jiang, Y. Fu, ``Fashion Style Generator'', In Proceedings of the Twenty-Sixth International Joint Conference on Artificial Intelligence. IJCAI, 2017, pp. 3721¨C3727.

\bibitem{SZhu}
S. Zhu, S. Fidler, R. Urtasun, D. Lin, C. Change, ``Be Your Own Prada: Fashion Synthesis with Structural Coherence'', arXiv preprint arXiv:1710.07346, 2017, pp. 1680¨C1688.

\bibitem{TAlotaiby}
R.Gravina, G.Fortino, ``Automatic methods for the detection of accelerative cardiac defense response'', \emph{IEEE Transactions on Affective Computing}, vol. 6, no. 3, 2016, pp. 286-298.

\bibitem{FPinel}
F. Pinel, ``What's Cooking with Chef Watson? An Interview with Lav Varshney and James Briscione'', \emph{IEEE Pervasive Computing}, vol. 14, no. 4, 2015, pp. 58-62.

\bibitem{BAStanescu}
B.A. Stanescu, C. Sandescu, B. M. Calapod, ``Automation Model for the Process of Creating Visual Identities in Educational Environments'', The International Scientific Conference eLearning and Software for Education, " Carol I" National Defence University, vol. 2, 2018, pp. 355-360.

\bibitem{YLeCun}
Y. LeCun, Y. Bengio, G. Hinton, ``Deep learning'', \emph{nature}, vol. 521, no. 7553, 2015, pp. 436.

\bibitem{AFink}
A. Fink, B. Graif, A. Neubauer. ``Brain correlates underlying creative thinking: EEG alpha activity in professional vs. novice dancers'', \emph{Neuroimage}, vol. 46, no. 3, 2009, pp. 854-862.


\bibitem{WTFreeman}
G.Fortino, S.Galzarano, R.Gravina, et al. ``A framework for collaborative computing and multi-sensor data fusion in body sensor networks'', \emph{Information Fusion}, vol.22, 2015, pp. 50-70.

\bibitem{RGravina}
R.Gravina, P.Alinia, H.Ghasemzadeh, et al. ``Multi-sensor fusion in body sensor networks: State-of-the-art and research challenges'', \emph{Information Fusion}, vol.35, 2017, pp. 68-80.

\bibitem{GFortino}
G.Fortino, R.Giannantonio, R.Gravina, et al. ``Enabling effective programming and flexible management of efficient body sensor network applications'', \emph{IEEE Transactions on Human-Machine Systems}, vol.43, no. 1, 2013, pp. 115-133.

\bibitem{AJChampandard}
A.J. Champandard, "Semantic Style Transfer and Turning Two-Bit Doodles into Fine Artworks", arXiv preprint arXiv:1603.01768, 2016.

\bibitem{ADietrich}
A. Dietrich, R. Kanso, ``A review of EEG, ERP, and neuroimaging studies of creativity and insight'', \emph{Psychological Bulletin}, vol. 136, no. 5, 2010, pp. 822-848.

\bibitem{MChen2017}
M. Chen, Y. Ma, Y. Li, D. Wu, Y. Zhang, Y. Chan, ``Wearable 2.0: Enabling human-cloud integration in next generation healthcare systems'', \emph{IEEE Communications Magazine}, vol. 55, no, 1, 2017, pp. 54-61.

\bibitem{LAGatys}
L.A. Gatys, A.S. Ecker, M. Bethge, ``A neural algorithm of artistic style'', \emph{IEEE Wireless Communications}, arXiv preprint arXiv:1508.06576, 2015.






\end{thebibliography}

\end{document}